\documentclass[letterpaper, 10pt, conference]{ieeeconf}
\IEEEoverridecommandlockouts  

\renewcommand{\baselinestretch}{0.96}
\usepackage{cite}
\usepackage{amsmath,amssymb,amsfonts}
\usepackage{algorithmic}
\usepackage{graphicx}
\usepackage{textcomp}
\usepackage{hyperref}
\usepackage{xcolor}
\usepackage[ruled,vlined]{algorithm2e}
\usepackage[capitalize]{cleveref}
\usepackage{tikz}
\usetikzlibrary{arrows.meta, positioning}

\def\BibTeX{{\rm B\kern-.05em{\sc i\kern-.025em b}\kern-.08em
    T\kern-.1667em\lower.7ex\hbox{E}\kern-.125emX}}

\title{\LARGE \bf
Wise Goose Chase: A Predictive Path Planning Algorithm for Dynamic Rebalancing in Ride-Hailing Systems
}
\author{Avalpreet Singh Brar$^{1}$, Rong Su$^{1}$, Christos G. Cassandras$^{2}$, Gioele Zardini$^{3}$  
\thanks{$^{1}$School of Electrical and Electronic Engineering, Nanyang Technological University, Singapore. (brar0002@e.ntu.edu.sg, rsu@ntu.edu.sg).}
\thanks{$^{2}$Division of Systems Engineering, Boston University, Brookline, MA 02446 USA (cgc@bu.edu)}
\thanks{$^{3}$Laboratory for Information and Decision Systems, Massachusetts Institute of Technology, Cambridge, MA, USA. (gzardini@mit.edu).}
}

\begin{document}

\maketitle

\begin{abstract}
Traditional rebalancing methods in ride-hailing systems direct idle drivers to fixed destinations, overlooking the fact that ride allocations frequently occur while cruising.
This destination-centric view fails to exploit the path-dependent nature of modern platforms, where real-time matching depends on the entire trajectory rather than a static endpoint.
We propose the Wise Goose Chase (WGC) algorithm, an event-triggered, driver-specific path planning framework that anticipates future matching opportunities by forecasting spatio-temporal supply and demand dynamics.
WGC uses a system of Retarded Functional Differential Equations (RFDEs) to model the evolution of idle driver density and passenger queues at the road-segment level, incorporating both en-route matching and competition among drivers.
Upon request, WGC computes personalized cruising paths that minimize each driver's expected time to allocation.
Monte Carlo simulations on synthetic urban networks show that WGC consistently outperforms baseline strategies, highlighting the advantage of predictive, context-aware rebalancing in dynamic mobility systems.
\end{abstract}

\section{Introduction}

In ride-hailing systems such as Uber and Lyft, taxi drivers operate as independent contractors competing to secure ride requests as quickly as possible. 
Once a driver becomes idle after a drop-off, they typically rely on personal heuristics, either \emph{hunting} (cruising for rides) or \emph{waiting} (staying stationary), to locate their next passenger~\cite{chen2021hunting}.
This decentralized behavior results in uncoordinated movement and persistent spatio-temporal mismatches between supply and demand.

To mitigate these inefficiencies, vehicle rebalancing has emerged as a key strategy.
Seminal work in~\cite{pavone2012robotic} introduced a fluid-based rebalancing framework, while~\cite{wallar2018vehicle} proposed agent-level repositioning policies that suggest destinations to idle drivers, improving their chances of being matched.
More recent efforts, such as~\cite{brar2020ensuring}, have incorporated fairness into rebalancing to ensure equitable treatment among drivers.
Incentive-based methods, including~\cite{sadeghi2019re} and \cite{ong2021driver}, dynamically offer monetary rewards to guide drivers toward high-demand areas.
Several works have further considered behavioral and operational complexities~\cite{ZardiniAnnRev2022}.
\cite{brar2024vehiclerebalancingadherenceuncertainty} modeled drivers' evolving adherence to recommendations based on prior experiences, while~\cite{chen2024rebalance} emphasized the influence of individual preferences on rebalancing efficacy.
In electric vehicles contexts,~\cite{guo2023vehicle} and~\cite{brar2022supply} incorporated battery charging constraints, and real-time deployment challenges have been tackled by~\cite{brar2021dynamic}.

However, most existing methods remain largely \emph{node-based}, recommending static destinations rather than considering en-route opportunities.
This design reflects legacy, station-based taxi models where pickups occurred at fixed points.
Yet, modern ride-hailing systems enable \emph{dynamic matching}, including while the driver is actively cruising.
In such contexts, static destination suggestions can be suboptimal.
A more adaptive approach involves recommending full \emph{routes} that maximize the likelihood of timely passenger allocation by accounting for evolving demand and supply along the way.
Furthermore, current rebalancing models typically adopt a fleet-level optimization perspective, periodically broadcasting decisions to all idle drivers to improve global metrics such as wait times or utilization.
While effective at scale, this can neglect individual context and responsiveness.
There is a growing need for \emph{event-driven, driver-initiated} systems that provide route suggestions on demand, triggered by the driver (e.g., via an in-app button).
Such systems can tailor recommendations to real-time, localized conditions while minimizing driver overload and reducing platform-wide computational burdens.

To this end, we propose the \textit{Wise Goose Chase} (WGC) model, a personalized, dynamic path recommendation framework.
Unlike traditional node-based methods, WGC forecasts supply and demand at the road-segment (edge) level and evaluates potential cruising paths based on expected time to allocation.
Recommendations are issued upon driven request and are adapted to both the driver's current location and the prevailing network state.

While path-based rebalancing remains underexplored, some related works exists.
For instance,~\cite{garg2018route} introduced a Monte Carlo Tree Search-based method (MDM) which suggests cruising routes based on historical pickup probabilities, updated via bandit feedback.
However, MDM is fundamentally reactive and time-discretized.
In contrast, WGC is anticipatory and model-based: it leverages retarded functional differential equations to forecast continuous-time supply-demand evolution at the edge level.
This enables the computation of \emph{survival probabilities}, the likelihood a driver remains unallocated while cruising a given path.
Unlike existing methods, WGC explicitly accountns for both en-route matching and spatial competition among drivers, providing a more realistic and strategic rebalancing mechanism.

\paragraph*{Statement of contribution}
This paper introduces the WGC algorithm, a novel, driver-specific path planning framework for dynamic rebalancing in ride-hailing systems.
WGC departs from conventional node-based rebalancing by operating at the road-segment level, capturing the spatio-temporal evolution of supply and demand via a system of RFDEs.
By forecasting future network conditions, WGC provides anticipatory, personalized cruising recommendations that minimize each driver's expected time to allocation.
Crucially, it models en-route matching opportunities and spatial competition among idle drivers, offering a more realistic, context-aware alternative to static destination-based strategies.
Through extensive numerical experiments, we demonstrate that WGC significantly outperforms baseline methods across a range of fleet sizes and demand scenarios, establishing the practical value of predictive, event-driven decision-making in modern mobility platforms.

\section{Wise Goose Chase (WGC) Model and Dynamics}
We consider a road network modeled as a directed graph \(G(\mathcal{V},\mathcal{E})\) as shown in \cref{fig:WGC_dynamics}, where \(\mathcal{V}\)  is the set of nodes (e.g., intersections) and \(\mathcal{E} \subset \mathcal{V} \times \mathcal{V}\)  is the set of directed edges (e.g., road segments). 
A fleet of \(N\) drivers operates on this network, transporting passengers between locations. 
Idle taxi drivers cruise the network according to a Continuous-Time Markov Chain (CTMC) process. 
While traversing an edge \(e=(u,v)\), a driver moves at a constant speed, taking \(\tau_e\) units of time to travel from node \(u\) to \(v\). 
Upon reaching node \( v \in V \), the driver selects a successor node \( w \in G^+(v)\) with a transition probability \( Q_{vw} \), where \(\sum_{w \in G^+(v)}Q_{wv} = 1\), as defined by the CTMC generator matrix \( Q \). Occupied drivers follow fixed paths from their current location on an edge \(e \in \mathcal{E}\) to a destination node \(u \in \mathcal{V}\), determined probabilistically. 
For an allocation at the edge \(e \in \mathcal{E}\), the probability that the driver will terminate the trip at node \(u \in \mathcal{V}\) is defined as \(R_{e \to u}\), where \(\sum_{u \in \mathcal{V}} R_{e \to u}=1\), and the probability \(R_{e \to u}\) is influenced by the popularity of destination \(u \in \mathcal{V}\). 
At any time \(t\in \mathbb{R_+}\), a driver can be in one of the following states: i) cruising on an edge while idle, ii) cruising on an edge while occupied (i.e., transporting a passenger), iii) waiting at a node while idle, or iv) waiting at a node while occupied. 
An idle driver may be matched with a passenger at any point along an edge \(e \in \mathcal{E}\).
Upon allocation, the driver transitions to the occupied state and proceeds toward a destination node \(u\in V\), arriving at time \(t' = t +\tau_{eu}\), where $\tau_{eu}$ is the travel time along the shortest path from the current location to $u$.
Passenger pick-up requests are generated along each edge \(e \in \mathcal{E}\) according to a non-homogeneous Poisson process with time-varying intensity. 
Each passenger has a finite tolerance for waiting, modeled as an exponential random variable with rate $\mu$.
If not matched with a driver before their patience expires, the passenger abandons the system.
\subsection{WGC State Variables}
Based on the setting above, we define the WGC state variables:

\begin{enumerate}
    \item \(D_e(t) \in \mathbb{R}_+\): Total number of idle drivers cruising on edge \(e \in \mathcal{E}\) at time \(t \in \mathbb{R}_+\).
    \item \(\tilde{D}_e(t) \in \mathbb{R}_+\): Total number of occupied drivers cruising on edge \(e \in \mathcal{E}\) at time \(t \in \mathbb{R}_+\).
    \item \(Q_e(t) \in \mathbb{R}_+\): Total number of passengers waiting on edge \(e \in \mathcal{E}\) at time \(t \in \mathbb{R}_+\).
    \item \(P_u(t) \in \mathbb{R}_+\): Total number of idle drivers at node \(u \in \mathcal{V}\) at time \(t \in \mathbb{R}_+\).
    \item \(\tilde{P}_u(t) \in \mathbb{R}_+\): Total number of occupied drivers at node \(u \in \mathcal{V}\) at time \(t \in \mathbb{R}_+\).
\end{enumerate}
Considering there are \(N\) drivers in our model, at any time \(t \in \mathbb{R}_+\) the following holds:
\begin{equation}\label{eqn:mass_conservation}
\sum_{e \in \mathcal{E}} D_e(t) +  \sum_{e \in \mathcal{E}} \tilde{D}_e(t) + \sum_{u \in \mathcal{V}} P_u(t) + \sum_{u \in \mathcal{V}} \tilde{P}_u(t) = N.
\end{equation}

\begin{figure*}[tb]
\centering

\begin{tikzpicture}[
  xscale=1.2, yscale=1.4,
  box/.style={draw, rectangle, minimum width=0.3cm, minimum height=0.4cm, inner sep=1pt, anchor=center, fill=black},
  dashed_box/.style={draw, dashed, rectangle, minimum width=0.3cm, minimum height=0.4cm, inner sep=1pt, anchor=center},
  line/.style={draw, -{Stealth[]}, thick}
  ]

\draw [dashed] (0.5,5) rectangle (14.5,8.6);

\node at (8,8.25) {\Large \textbf{Wise Goose Chase (WGC) System Dynamics}};

\node[anchor=north west] at (1,8) {\normalsize $\mathbf{Edge\text{-}Level\ Dynamics\ of\ Waiting\ Passengers:}$};
\node[anchor=north west] at (1,7.75) {\normalsize $\frac{dQ_{e}(t)}{dt} = \lambda_{e}(t) - \mu Q_{e}(t) - A_{e}(t)$};

\node[anchor=north west] at (1, 7) {\normalsize $\mathbf{Edge\text{-}Level\ Dynamics\ of\ Idle\ Drivers:}$};
\node[anchor=north west] at (1, 6.75) {\normalsize $\frac{dD_{e}(t)}{dt} = P_{u}(t) \cdot Q_{uv} - A_{e}(t) - P_{u}(t-\tau_{e}) \cdot Q_{uv} \cdot G_{uv}(t)$};

\node[anchor=north west] at (1, 6) {\normalsize $\mathbf{Node\text{-}Level\ Dynamics\ of\ Idle\ Drivers:}$};
\node[anchor=north west] at (1, 5.75) {\normalsize $\frac{dP_{u}(t)}{dt} = \sum_{i} R_{i\rightarrow u} \cdot A_{i}(t-\tau_{iu}) - \sum_{v\in N^{+}(u)} P_{u}(t) \cdot Q_{uv} + \sum_{w\in N^{-}(u)} P_{w}(t-\tau_{wu}) \cdot Q_{wu} \cdot G_{wu}(t)$};

\node at (4, 4.5) {\normalsize $\mathbf{Initial\ Condition}$};
\node at (10,4.5) {\normalsize $\mathbf{Predicted\ Trajectories}$};

\node at (4, 2.5) {\includegraphics[width=150pt]{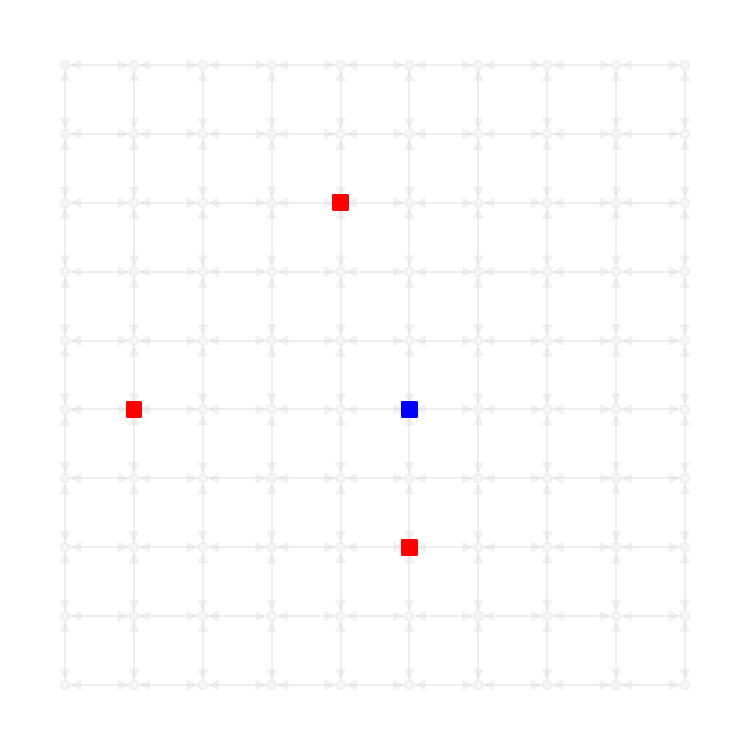}};
\node at (10,2.5) {\includegraphics[width=150pt]{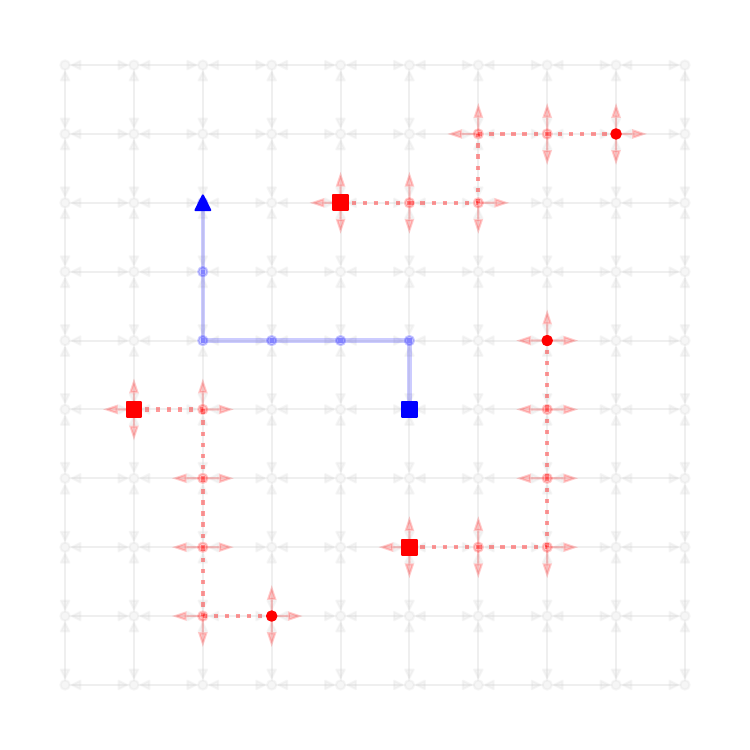}};

\end{tikzpicture}

\caption{
This figure provides an overview of the WGC system dynamics.  
The top part presents the set of coupled differential equations describing the evolution of unmatched passengers at the edge level, idle drivers cruising on edges, and idle drivers located at nodes.  
The bottom part illustrates an example of the predicted state trajectories, showing the evolution from the initial distribution of drivers and passengers to the forecasted supply-demand patterns over time.
}
\label{fig:WGC_dynamics}
\end{figure*}
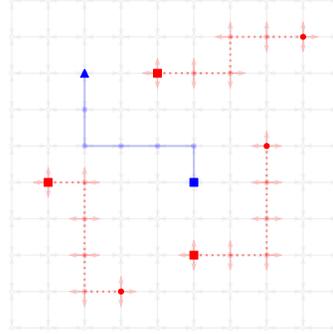

To study the evolution of the WGC system, we derive a system of RFDEs as shown in \cref{fig:WGC_dynamics} that describe the time evolution of the total number of unmatched passengers \( Q_e(t) \), idle drivers, \( D_e(t) \), occupied drivers \( \tilde{D}_e(t) \) on edge \(e \in \mathcal{E}\), and idle drivers \(P_u(t)\) at node \(u \in \mathcal{V}\) at time \(t \in \mathbb{R}_+\).

\begin{table}[t]
\centering
\small
\caption{Summary of WGC Variables.}
\renewcommand{\arraystretch}{1.3}
\begin{tabular}{|p{1cm}|p{6.5cm}|}
\hline
\textbf{Symbol} & \textbf{Description} \\
\hline
\( D_e(t) \) & Number of idle drivers on edge \( e \) at time \( t \) \\
\( \tilde{D}_e(t) \) & Number of occupied drivers on edge \( e \) at time \( t \) \\
\( Q_e(t) \) & Number of passengers on edge \( e \) at time \( t \) \\
\( P_u(t) \) & Number of idle drivers at node \( u \) at time \( t \) \\
\( \tilde{P}_u(t) \) & Number of occupied drivers at node \( u \) at time \( t \) \\
\( A_e(t) \) & Number of drivers allocated on edge \( e \) at time \( t \) \\
\( Q_{uv} \) & Transition rate from node \( u \) to node \( v \) in the CTMC \\
\( \tau_e \) & Time required by cruising driver to cross edge \( e \) \\
\( \tau_{eu} \) & Time occupied driver takes to go from edge \(e \to u\) \\
\( R_{e \to u} \) & Probability of trip to go from edge \(e \to u\) \\
\hline
\end{tabular}
\label{tab:notation-idle}
\end{table}

\subsection{Passenger Queue Dynamics}

Let \( \lambda_e(t) \) denote the time-varying rate of non-homogeneous Poisson process with which the passenger pick-up requests arrive at edge \(e \in \mathcal{E}\). Arriving requests queue up in parallel servers, and there is no priority based on the waiting time. 

Each arriving passenger is assumed to have a finite \textit{patience time} \( T_p \), which represents the maximum duration they are willing to wait to be picked up before abandoning the system. We model \( T_p \) as an exponential random variable with rate parameter \( \mu > 0 \). That is, for any \( t \geq 0 \), the probability that a passenger is willing to wait at least \( t \) units of time is given by the following equation:
\begin{equation}
P(T_p > t) = \int_t^{\infty} \mu e^{-\mu x} \, dx = e^{-\mu t}
\end{equation}

Using the memoryless property of the exponential distribution, it is easy to show that:
\begin{equation}
\begin{aligned}
P(T_p \in [t, t + \Delta t) \mid T_p > t) 
&= \frac{P(T_p \in [t, t + \Delta t))}{P(T_p > t)}  = \mu \Delta t.
\end{aligned}
\end{equation}

Now consider \( Q_e(t) \) unmatched passengers on edge \( e \) at time \( t \). 
Each of them independently follow their own exponential abandonment clock. 
Let us define the indicator variable \( X_i(\Delta t) \) for passenger \( i \in \{1, 2, \dots, Q_e(t)\} \) as:
\begin{equation}
X_i(\Delta t) = 
\begin{cases}
1, & \text{if passenger } i \text{ abandons in } [t, t + \Delta t), \\
0, & \text{otherwise}.
\end{cases}
\end{equation}

Then, using the result above:
\begin{equation}
\mathbb{E}[X_i(\Delta t)] = P(\text{passenger \(i\) abandons in } [t, t + \Delta t)) = \mu \Delta t
\end{equation}

So, the total number of abandonments in \( [t, t + \Delta t) \) be:
\begin{equation}
\alpha(\Delta t) = \sum_{i=1}^{Q_e(t)} X_i(\Delta t).
\end{equation}

Taking expectation, one has:
\begin{equation}
\mathbb{E}[\alpha(\Delta t)] = \sum_{i=1}^{Q_e(t)} \mathbb{E}[X_i(\Delta t)] = Q_e(t) \cdot \mu \Delta t.
\end{equation}

To obtain the abandonment \emph{rate}, we compute the limit:
\begin{equation}
\lim_{\Delta t \to 0} \frac{\mathbb{E}[\alpha(\Delta t)]}{\Delta t} 
= \lim_{\Delta t \to 0} \frac{Q_e(t) \cdot \mu \Delta t}{\Delta t} 
= \mu Q_e(t).
\end{equation}

Thus, the expected number of passengers abandoning the system per unit time at time \( t \) is given by:
\begin{equation}
\text{Abandonment rate at edge \(e\)} = \mu Q_e(t).
\end{equation}

This yields a linear depletion term in the queue dynamics due to passenger impatience.
In addition to abandonment, passengers may leave the queue upon being matched with an idle driver. 
Let \( D_e(t) \) denote the number of idle drivers on edge \( e \) at time \( t \). We assume that whenever both an idle driver and a waiting passenger are present, a match occurs immediately. 
Thus, the instantaneous matching rate is given by:
\begin{equation}
A_e(t) = \min(D_e(t), Q_e(t)).
\end{equation}

Combining all components, the net rate of change of the number of unmatched passengers on edge \( e \in \mathcal{E} \) is given by the following differential equation:
\begin{equation}
\label{eqn:passenger-ode}
\frac{d Q_e(t)}{dt} = \lambda_e(t) - A_e(t) - \mu Q_e(t),
\end{equation}
where:
\begin{itemize}
    \item \( \lambda_e(t) \) is the rate at which the passenger requests arrive;
    \item \( A_e(t) \) is the rate at which the passengers are matched with idle drivers;
    \item \( \mu Q_e(t) \) is the rate at which the passengers abandon the system due to impatience.
\end{itemize}

\subsection{Driver Dynamics}
 As specified in \cref{eqn:mass_conservation} the total number of drivers in the system remains constant over time. 
 However, to compute an optimal route for a tagged driver, it is essential to account for the presence of other idle drivers along the recommended path.
 Such idle drivers represent local competition, reducing the tagged driver's likelihood of securing a ride.
 To capture the spatio-temporal dynamics of this competition, we model the evolution of idle driver density over the network \(G(E,V)\) using the follosing system of RFDEs.

\subsubsection{Edge Level Idle Driver Dynamics}
Idle drivers cruise on an edge \(e = (u,v)\) for a maximum duration of \(\tau_e\) time units. 
During this time, the drivers may get allocated and the status of the driver is changed to occupied. 
Drivers that remain unallocated exit the edge on which they were cruising and enter the downstream node \( v \). 
At node \( v \), drivers are partitioned into outgoing edges \(\{ e' = (v, w) : w \in \mathcal{N}^+(v) \}\) according to transition probabilities specified by the CTMC transition matrix \( Q \).
Here, for any node \( u \in \mathcal{V} \), we define the set of downstream neighbors as \(\mathcal{N}^+(u) = \{ w \in \mathcal{V} : (u, w) \in \mathcal{E} \}\) and the set of upstream neighbors as \(\mathcal{N}^-(u) = \{ v \in \mathcal{V} : (v, u) \in \mathcal{E} \}\). 
This way, at each time \(t \in \mathbb{R}_+\) the total number of idle drivers in an edge keeps evolving due to the inflow of idle drivers at the upstream node \(u\), outflow of the drivers at the downstream node \(v\), and the allocation of the drivers while cruising on the road.  
Next, we derive the expression that captures the dynamics of the total number of idle drivers on the edge considering the above-mentioned three factors:

\begin{enumerate}
    \item \textbf{Inflow from node to edge:} The total number of drivers standing idle at the node \(u \in \mathcal{V}\) at time \(t \in \mathbb{R}_+\), i.e., \(P_u(t)\) gets divided proportionally to the downstream edges at the rate \( P_u(t) \cdot Q_{uv} \). This is defined as the inflow into the edge \(e \in \mathcal{E}\) at time \(t \in \mathbb{R}_+\).

    \item \textbf{Outflow due to traversal:} Outflow at time \(t\) consists of a fraction of \(P_u(t-\tau_e) \cdot Q_{uv}\) drivers that entered the edge \( e = (u,v) \) at time \( t - \tau_e \), and reach the downstream node \( v \) at time \(t\). These drivers complete the \(\tau_e\) time unit traversal on the edge \(e\) without getting allocated. 
    
    To determine the outflow, we define the concept of survival probability. The survival probability (i.e., the probability that an idle driver cruising on \( e \) is not matched between time \( t - \tau_e \) and \( t \)) is derived as follows.

    Assuming \( D_e(s) > 0 \), we define the \emph{instantaneous allocation hazard rate} at time \( s \) as:
    \begin{equation}
      h_e(s) = \frac{A_e(s)}{D_e(s)}  
    \end{equation}
    
    Now, consider a small time interval \( [s, s + \Delta s] \subset [t - \tau_e, t] \). Let \( D_e(s) \) denote the number of idle drivers cruising on edge \( e \) at time \( s \), and let \( A_e(s) \) be the number of driver-passenger matches (i.e., allocations) that occur on edge \( e \) per unit time. We make the assumption that each idle driver on edge \( e \) is equally likely to be matched. Under this assumption, the probability that any particular driver is matched in the small interval \( [s, s + \Delta s] \) is approximately:
    \begin{equation}
        P(\text{allocated in } [s, s+ \Delta s]) = \frac{A_e(s)}{D_e(s)} \cdot \Delta s.
    \end{equation}
   
    Thus, the probability that the driver \emph{survives unmatched} (i.e., is not allocated) during this time interval is:
    \begin{equation}
    P(\text{survive in } [s, s+ \Delta s]) = 1 - \frac{A_e(s)}{D_e(s)} \cdot \Delta s.
    \end{equation}
    
    To survive unmatched over the entire interval \( [t - \tau_e, t] \), the driver must survive each small subinterval. Therefore, the survival probability is given by the product:
    \begin{equation}
     P(\text{survive in } [t - \tau_e, t]) = \prod_{s = t - \tau_e}^{t} \left(1 - h_e(s) \cdot \Delta s \right). 
    \end{equation}
    
    Taking the logarithm of the product:
    \begin{equation}
        \begin{aligned}
            \log(P(\text{survive in } & [t - \tau_e, t]) = \\
            &\sum_{s = t - \tau_e}^{t} \log\left(1 - h_e(s) \cdot \Delta s \right)
        \end{aligned}
    \end{equation}
    
    Using the approximation \( \log(1 - x) \approx -x \) for small \( x \), we have:
    \begin{equation}
    \sum_{s = t - \tau_e}^{t} \log\left(1 - h_e(s) \cdot \Delta s \right) \approx -\sum_{s = t - \tau_e}^{t} h_e(s) \cdot \Delta s.  
    \end{equation}
    
    As \( \Delta s \to 0 \), this Riemann sum converges to an integral:
    \begin{equation}
    \log(P(\text{survive in } [t - \tau_e, t]) \to - \int_{t - \tau_e}^{t} h_e(s)\, ds.
    \end{equation}
    
    Taking the exponential of both sides gives the survival probability:
    \begin{equation}\label{eqn:survival probability}
    P(\text{survive in } [t - \tau_e, t] = \exp\left( - \int_{t - \tau_e}^{t} \frac{A_e(s)}{D_e(s)} \, ds \right).
    \end{equation} 
    
    We define the following operator to represent the survival probability for the driver traversing the edge \(e = (u,v)\) from \([t-\tau_e, t]\):
    \begin{equation}
        G_{uv}(t) = \exp\left( - \int_{t - \tau_e}^{t} \frac{A_e(s)}{D_e(s)} \, ds \right).
    \end{equation}
    
    Finally, suppose the number of idle drivers at node \( u \) at time \( t - \tau_e \) is \( P_u(t - \tau_e) \), and the probability of transitioning from node \( u \) to \( v \) is \( Q_{uv} \). Then, the number of drivers who enter edge \( e = (u,v) \) at time \( t - \tau_e \) is:
    \begin{equation}
    P_u(t - \tau_e) \cdot Q_{uv}.
    \end{equation}
    
    Multiplying by the survival probability gives the number of unmatched drivers who exit edge \( e \) at time \( t \):
    \begin{equation}
    P_u(t - \tau_e) \cdot Q_{uv} \cdot G_{uv}(t).
    \end{equation}

    \item \textbf{Matching while cruising:} Drivers cruising on an edge \(e \in \mathcal{E}\) are immediately removed from the pool of idle drivers when allocated to the passengers waiting on the edge. Thus at time \(t \in \mathbb{R}_+\), the drivers are removed at a rate which is same as the allocation rate \( A_e(t) \).
\end{enumerate}

Combining the above three sources, we define the dynamics of the number of idle drivers on edge \(e \in \mathcal{E}\) using the following Retarded Differential Equations for edge \( e = (u,v) \), which is given as follows:
\begin{equation}
\label{eq:idle-driver-ode}
\begin{aligned}
\frac{d D_e(t)}{dt} = P_u(t) \cdot Q_{uv} &- A_e(t)\\
&- P_u(t - \tau_e) \cdot Q_{uv} \cdot G_{uv}(t).
\end{aligned}
\end{equation}

\subsubsection{Node-Level Idle Driver Dynamics}
There are three sources that influence the number of drivers at a node \(u \in \mathcal{V}\): 
\begin{enumerate}
    \item \textbf{Inflow of idle drivers:} The idle drivers from the upstream nodes \(w \in \mathcal{N}^-(u)\) 
     of the target node \(u \in \mathcal{V}\) that started traveling \(\tau_{wu}\) time units ago reach the target node \(u \in \mathcal{V}\) at time \(t \in \mathbb{R}_+\) after surviving the allocation process on the edges:
     \begin{equation}
     \sum_{w \in \mathcal{N}^-(u)} P_w(t - \tau_{wu}) \cdot Q_{wu} \cdot G_{wu}(t).
     \end{equation}
    \item \textbf{Return of occupied drivers:} When a driver gets allocated on edge \(e \in \mathcal{E}\) and travels from edge \(e \in E\) to one of the nodes \(u \in \mathcal{V}\) where the trip ends. It takes a driver \(\tau_{eu}\) time to travel from edge \(e \in  \mathcal{E}\) to node \(u \in \mathcal{V}\). Then \(A_e(t-\tau_{eu})\) drivers who were allocated on edge \(e \in \mathcal{E} \) at time \(t-\tau_{eu}\), will end their trip at node \(u \in \mathcal{V}\) at time \(t \in \mathbb{R}_+\), so the total number of returning occupied drivers is as follows:
    \begin{equation}
    \sum_{e \in \mathcal{E}} R_{e \to u} \cdot A_e(t - \tau_{eu}).
    \end{equation}
    \item \textbf{Outflow of idle drivers:} the idle drivers currently on node \(u \in \mathcal{V}\) get distributed to head towards the downstream nodes \(w \in \mathcal{N}^+(u)\) leading to driver outflow:
    \begin{equation}
    \sum_{v \in \mathcal{N}^+(u)} P_u(t) \cdot Q_{uv}.
    \end{equation}
Including the above mentioned factors, the dynamics of the idle drivers at the node \(u \in \mathcal{V}\) evolve as follows:
\end{enumerate}
\begin{equation}
\label{eq:idle-node-dynamics}
\begin{aligned}
\frac{d P_u(t)}{dt} & = \sum_{e \in \mathcal{E}} R_{e \to u} \cdot A_e(t - \tau_{eu}) - \sum_{v \in \mathcal{N}^+(u)} P_u(t) \cdot Q_{uv} \\
& + \sum_{w \in \mathcal{N}^-(u)} P_w(t - \tau_{wu}) \cdot Q_{wu} \cdot G_{wu}(t).
\end{aligned}
\end{equation}

\subsection{Optimal Path Planning for Fastest Allocation}

We consider the decision problem faced by a tagged idle driver initially located at node \( v_0 \in \mathcal{V} \) at time \( t = 0 \), as illustrated on the left side of \cref{fig:WGC_dynamics}.  
The objective is to select a sequence of road segments (edges) to traverse to minimize the expected time until the connection is made with a passenger.
Let \( \pi = (e_1, e_2, \dots, e_K) \) denote a finite path over the edge set \( \mathcal{E} \), where each edge \( e_k = (v_{k-1}, v_k) \), and the path begins at node \( v_0 \).  
Let \( \tau_{e_k} \) denote the idle travel time across edge \( e_k \), and define the cumulative traversal time up to the end of edge \( e_k \) as:
\begin{equation}
T_k = \sum_{j=1}^k \tau_{e_j}.
\end{equation}
The total planned cruising time along the path is:
\begin{equation}
T_\pi = \sum_{k=1}^K \tau_{e_k}.
\end{equation}

Let \( Q_e(t) \) and \( D_e(t) \) denote the predicted number of unmatched passengers and idle drivers, respectively, on edge \( e \) at time \( t \), as forecasted by the WGC model (solutions to the RFDEs described in \cref{fig:WGC_dynamics}).  
The instantaneous matching rate on edge \( e \) at time \( t \) is defined as:
\begin{equation}
A_e(t) = \min\left(Q_e(t), D_e(t)\right).
\end{equation}

Given the previously defined survival operator \( G_{uv}(t) \), which represents the probability of not being allocated when crossing the edge \( e = (u,v) \) during the interval \([t-\tau_e, t]\):
\begin{equation}
G_{uv}(t) = \exp\left( -\int_{t-\tau_e}^{t} \frac{A_e(s)}{D_e(s)} \, ds \right).
\end{equation}
We express the overall survival probability \( S(t) \) at time \( t \) while following the path \( \pi \) as the product of survival operators along the edges traversed up to time \( t \).
Specifically, if \( t \in [T_{k-1}, T_k) \), i.e., the driver is currently on edge \( e_k \), then:
\begin{equation}
S(t) = \prod_{j=1}^{k-1} G_{v_{j-1}v_j}(T_j) \cdot G_{v_{k-1}v_k}(t),
\end{equation}
where \( G_{v_{j-1}v_j}(T_j) \) accounts for survival over previously completed edges, and \( G_{v_{k-1}v_k}(t) \) accounts for survival up to the current time along the current edge.

Thus, the expected allocation time along path \( \pi \) is given by:
\begin{equation}\label{eqn:expected_allocation_time}
\mathbb{E}[T_{\text{alloc}} \mid \pi] = \int_0^{T_\pi} S(t) \, dt.
\end{equation}

The optimal path \( \pi^* \in \mathcal{P}_{v_0} \), where \( \mathcal{P}_{v_0} \) denotes the set of feasible paths originating from node \( v_0 \), is obtained by solving:
\begin{equation}
\pi^* = \arg\min_{\pi \in \mathcal{P}_{v_0}} \mathbb{E}[T_{\text{alloc}} \mid \pi].
\end{equation}

\subsection{Numerical Approximation}

The evolution of the WGC system is governed by a system of delay differential equations defined in continuous time. To enable simulation and evaluation of candidate paths, we discretize the system using the forward Euler method with a fixed step size \( \Delta t \). The state vector at time \( t \in [0, T] \) is given by:
\begin{equation}
y(t) = \left( Q_e(t),\ D_e(t),\ P_u(t) \right)_{e \in \mathcal{E},\ u \in \mathcal{V}}    
\end{equation}

Let \( t_k = k \Delta t \) for \( k = 0, 1, \dots, \lfloor T/\Delta t \rfloor \). The continuous dynamics are approximated at these discrete time points by updating the state variables according to the following rules:

\paragraph{Passenger Queues.}
The number of unmatched passengers on edge \( e \) evolves based on new arrivals, successful matches, and abandonment:
\begin{equation}\label{eqn:numerical_passenger_dynamics}
Q_e(t_{k+1}) = Q_e(t_k) + \Delta t \left( \lambda_e(t_k) - A_e(t_k) - \mu Q_e(t_k) \right).
\end{equation}

\paragraph{Idle Drivers on Edges.}
The number of idle drivers cruising on edge \( e = (u,v) \) evolves due to inflows from node \( u \), real-time allocations on edge \( e \), and outflows from drivers completing their traversal of \( e \):
\begin{equation}\label{eqn:numerical_edge_dynamics}
\begin{aligned}
D_e(t_{k+1}) & =\  D_e(t_k) + \Delta t \Bigg[
P_u(t_k) \cdot Q_{uv} - A_e(t_k) \\
& - P_u(t_k - \tau_e) Q_{uv} \cdot \tilde{G}_{uv}(\Delta t)
\Bigg],
\end{aligned}
\end{equation}
where \( m = \tau_e / \Delta t \) is the number of discrete steps across the edge travel duration and,
\begin{equation}
    \tilde{G}_{uv}(\Delta t) = \exp\left( - \sum_{j=0}^{m-1} \frac{A_e(t_k - j \Delta t)}{D_e(t_k - j \Delta t)} \cdot \Delta t \right)
\end{equation}

\paragraph{Idle Drivers at Nodes.}
The number of idle drivers at node \( u \) evolves based on drivers returning from completed trips, outflows to outgoing edges, and inflows from neighboring nodes completing edge traversals:
\begin{equation}\label{eqn:numerical_node_dynamics}
\begin{aligned}
P_u(t_{k+1}) & =\  P_u(t_k) + \Delta t \Bigg[\sum_{i \in \mathcal{E}} R_{i \to u} \cdot A_i(t_k - \tau_{iu}) \\
& - \sum_{v \in N^+(u)} P_u(t_k) \cdot Q_{uv} \\
& + \sum_{w \in N^-(u)} P_w(t_k - \tau_{wu}) \cdot Q_{wu} \cdot \tilde{G}_{wu}(\Delta t)
\Bigg],
\end{aligned}
\end{equation}
where \( m' = \tau_{wu} / \Delta t \) for each edge \( (w,u) \in \mathcal{E} \) and,
\begin{equation}
     \tilde{G}_{wu}(\Delta t) = \exp\left( - \sum_{j=0}^{m'-1} \frac{A_{wu}(t_k - j \Delta t)}{D_{wu}(t_k - j \Delta t)} \cdot \Delta t \right)
\end{equation}

\paragraph{Expected time to allocation.}
The expected time the driver needs to cruise before getting allocated along a candidate path \( \pi \) is approximated using discrete survival probability updates. Initialization is performed with survival probability \( S(0) = 1 \) and cumulative expected time \( T_{\text{alloc}}(0) = 0 \). For each \( t_k\), the updates are given by:
\begin{align} \label{eqn:numerical_expected_allocation_time}
&T_{\text{alloc}}(t_{k+1}) = T_{\text{alloc}}(t_k) + S(t_k) \Delta t, \\[6pt]
&S(t_{k+1}) = S(t_k) \cdot \exp\left( -h(t_k) \Delta t \right),
\end{align}
where \( h(t_k) \) is evaluated based on the current edge being traversed. The traversal across edges is tracked by accumulating elapsed travel time, and transitions occur once the edge travel time \( \tau_e \) is completed. Computation terminates early if \( S(t_k) \) falls below a small threshold \( \epsilon \).

\subsection{Algorithm: Wise Goose Chase (WGC)}

The WGC algorithm computes a personalized, anticipatory cruising path for an idle driver in a ride-hailing system.  
Starting from the driver's current location, WGC predicts the future evolution of unmatched passengers and idle drivers by numerically integrating the system dynamics given by \cref{eqn:numerical_passenger_dynamics}, \cref{eqn:numerical_edge_dynamics}, and \cref{eqn:numerical_node_dynamics}.  
Based on these forecasts, it evaluates candidate paths by simulating the driver's survival probability and estimating the expected allocation time along each path using \cref{eqn:numerical_expected_allocation_time}.  
The path that minimizes the expected time to passenger allocation is selected, thereby capturing both en-route matching opportunities and evolving supply-demand competition.

\begin{algorithm}[tb]
\caption{Wise Goose Chase (WGC)}
\KwIn{
Graph \( G = (\mathcal{V}, \mathcal{E}) \); \\
Initial states \( Q_e(0) \), \( P_u(0) \); \\
Simulation parameters: \( T \), \( \Delta t \), \( \mu \); \\
Starting node \( v_0 \); max. path length \( L \); threshold \( \epsilon \).
}
\KwOut{Optimal path \( \pi^* \)}
\vspace{2pt}

\textbf{Stage 1: Numerical Trajectory Generation}\;
Discretize time: \( t_k = k \Delta t \)\;
\For{each \( t_k \)}{
    Update \( Q_e(t_k) \), \( D_e(t_k) \), \( P_u(t_k) \) using \cref{eqn:numerical_passenger_dynamics}, \cref{eqn:numerical_edge_dynamics}, \cref{eqn:numerical_node_dynamics}\;
}

\vspace{2pt}
\textbf{Stage 2: Path Evaluation}\;
Initialize candidate set \( \Pi \gets \emptyset \)\;
\For{each simple path \( \pi = (e_1, \dots, e_K) \) from \( v_0 \) with \( K \leq L \)}{
    Set \( \mathbb{E}[T_{\text{alloc}} \mid \pi] \gets 0 \), \( S \gets 1 \), \( t \gets 0 \), \( e \gets e_1 \)\;
    Initialize edge index \( k \gets 1 \), elapsed time \( \tau_{\text{elap}} \gets 0 \)\;
    
    \While{still on path and \( S > \epsilon \)}{
        Set \( t_k = \lfloor t / \Delta t \rfloor \)\;
        Compute hazard \( h(t_k) = A_{e}(t_k) / D_{e}(t_k) \)\;
        Update \( \mathbb{E}[T_{\text{alloc}} \mid \pi] \mathrel{+}= S \Delta t \), \( S \gets S \exp(-h(t_k) \Delta t) \)\;
        Advance \( t \gets t + \Delta t \), \( \tau_{\text{elap}} \gets \tau_{\text{elap}} + \Delta t \)\;
        \If{\( \tau_{\text{elap}} \geq \tau_e \)}{
            Advance to next edge \( e_{k+1} \), reset \( \tau_{\text{elap}} \gets 0 \)\;
        }
    }
    Add \( (\pi, \mathbb{E}[T_{\text{alloc}} \mid \pi]) \) to \( \Pi \)\;
}
\Return \( \pi^* = \arg\min_{\pi \in \Pi} \mathbb{E}[T_{\text{alloc}} \mid \pi] \)\;
\end{algorithm}

\section{Experiments and Observations}

We evaluate the effectiveness of the proposed WGC-based routing strategy via Monte Carlo simulations on a synthetic urban environment modeled as a directed $10 \times 10$ grid network.
Each node represents an intersection, and each directed edge represents a road segment, allowing bidirectional movement through distinct node pairs.
Simulations are executed over a fixed time horizon of 600 seconds, discretized into uniform time steps. 
For each fleet size \( N \in \{100, 500, 1000, 2000, 4000, 5000\} \), 100 independent trials are conducted using randomized initial conditions to obtain statistically robust performance estimates.

At the beginning of each trial, idle drivers are randomly distributed across the nodes according to a normalized multinomial sampling process, ensuring non-negative driver counts summing to \( N \). 
Passenger arrivals on road segments follow independent and time-varying Poisson processes.
Spatial heterogeneity is introduced by designating a subset of edges as hotspots (assigned elevated arrival rates) and others as cold zones (with zero arrivals). 
The remaining edges receive base arrival rates sampled from a uniform distribution. A sinusoidal perturbation is added to all arrival rate profiles to emulate temporal fluctuations.
Idle drivers evolve according to a CTMC, selecting outgoing edges based on a transition probability matrix. 
The simulation tracks edge-level idle driver dynamics \( D_e(t) \), passenger queues \( Q_e(t) \), and matching rates \( A_e(t) \), updated using a forward Euler integration scheme. 
En-route passenger allocation is supported, and unmatched passengers abandon the system based on an exponential patience distribution with a mean of 10 seconds. 
Post-matching, occupied drivers are routed to destinations sampled from a global popularity distribution, and are assumed to re-enter the system after completing trips based on edge-to-node return delay distributions.

\subsection{Long-Term Behavior of the WGC System Dynamics}
\begin{figure}[tb]
    \centering
    \includegraphics[width=1\linewidth]{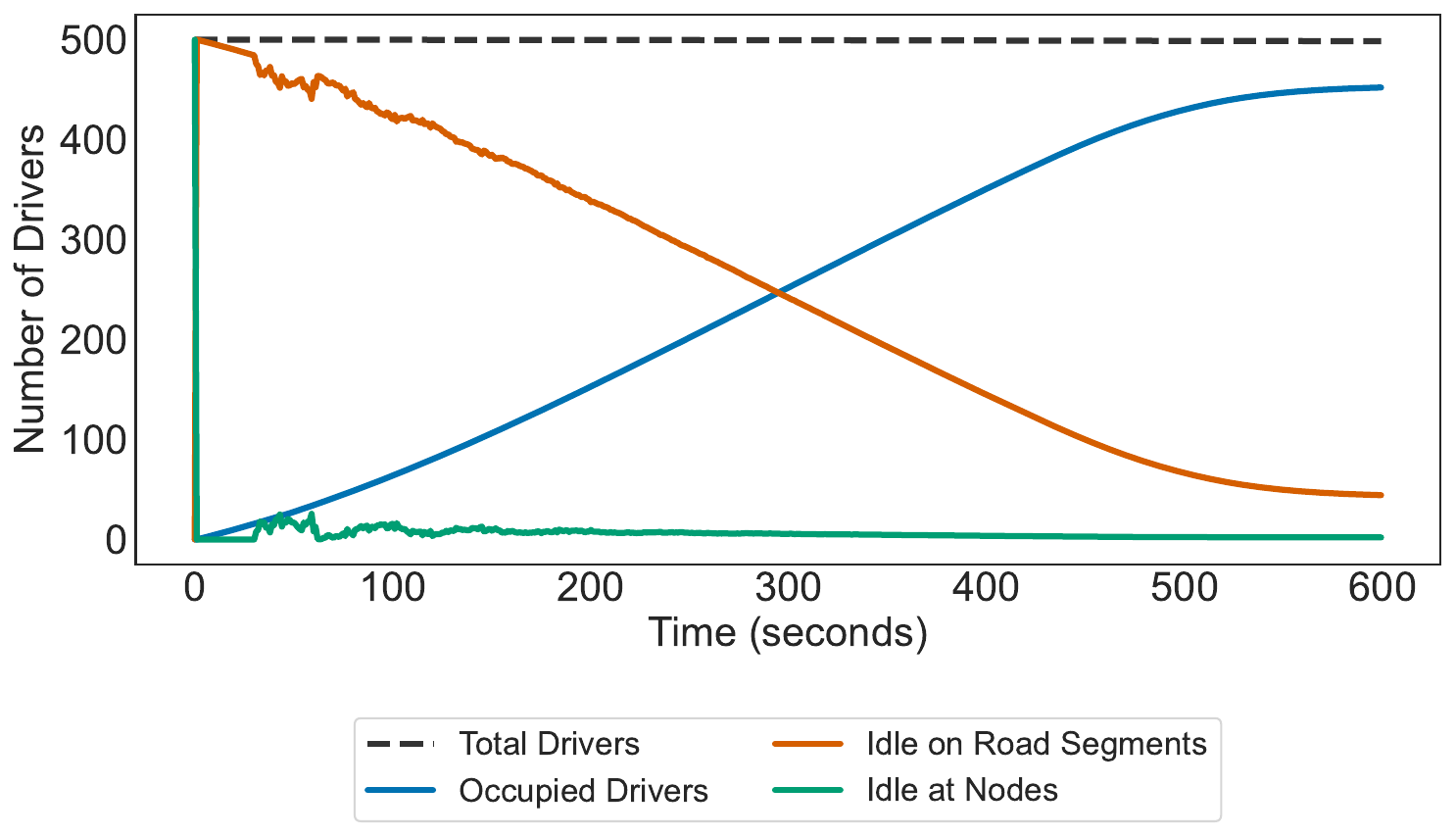}
    \caption{
    Convergence behavior of the WGC system dynamics over time.
    }
    \label{fig:dynamics}
\end{figure}

The temporal evolution of the WGC system is depicted in \cref{fig:dynamics}. 
At the beginning of the simulation, all drivers are idle and distributed across the road network. As the system evolves:
\begin{itemize}
    \item The number of \emph{occupied drivers} steadily increases as idle drivers are matched to incoming passenger requests.
    \item The number of \emph{idle drivers cruising on road segments} decreases as more drivers transition to occupied status.
    \item The number of \emph{idle drivers at nodes} remains consistently low, indicating minimal congestion or waiting at intersections.
    \item The \emph{total number of drivers} remains constant over time, verifying the conservation of driver mass within the system.
\end{itemize}

\subsection{Baseline Strategies}

To evaluate the effectiveness of the proposed WGC-based routing strategy, we compare its performance against the following three baseline policies:

\begin{enumerate}
    \item \textbf{Random Walk}: Under this strategy, each driver randomly selects an outgoing edge at all intersections. Drivers do not utilize any information on passenger demand or arrival rates, resulting in purely stochastic cruising behavior.
    
    \item \textbf{Greedy}: In this strategy, drivers select outgoing edges with the maximum passenger arrival rate. This policy captures a myopic demand-seeking behavior, ignoring future driver competition and broader system dynamics.

    \item \textbf{Hotspot Guided}: In this strategy, drivers are directed toward predefined hotspot regions characterized by historically high passenger demand. Drivers preferentially move toward these areas without dynamically optimizing for en-route matching or real-time competition, providing a destination based repositioning benchmark.
\end{enumerate}

These baselines span a range of operational behaviors, from naive random exploration to heuristic demand targeting. Comparative evaluation against these baselines highlights the benefits of anticipatory and system-aware path planning enabled by the WGC approach.

\subsection{Comparison of Strategy Performances}
The effectiveness of the proposed WGC-based path planning strategy is evaluated against three baseline strategies: Random Walk, Greedy, and Hotspot Guided. The comparative results are summarized in \cref{tab:strategy_comparison}.

\begin{figure}[tb]
    \centering
    \includegraphics[width=0.7\linewidth]{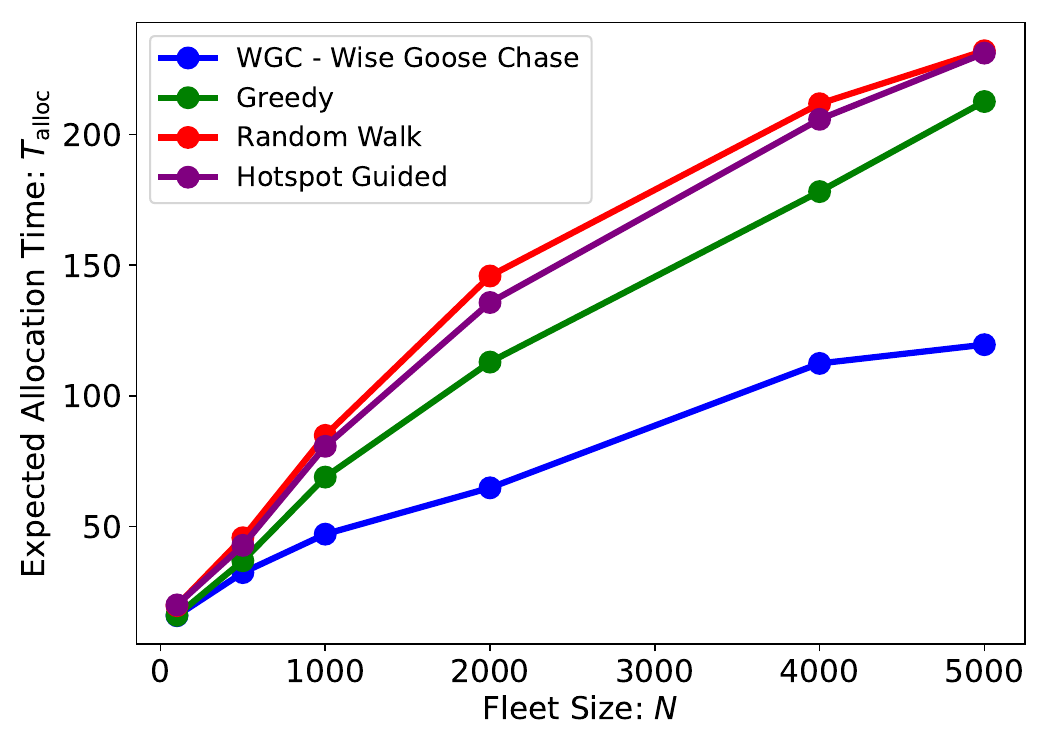}
    \caption{Comparison of expected allocation time (\(T_{alloc}\)) under different routing strategies across fleet sizes (\(N\)).}
    \label{fig:performance_comparison}
\end{figure}

As shown in~\cref{tab:strategy_comparison}, the \emph{WGC} strategy consistently achieves the lowest expected allocation time in all fleet sizes, outperforming all baseline methods. 
The \emph{Greedy} strategy, which selects routes based on local passenger arrival rates, offers competitive performance at smaller fleet sizes, but does not capture future competition effects, leading to suboptimal results as the fleet size grows.  
The \emph{Random Walk} strategy performs the worst due to its uninformed and inefficient cruising behavior. 
The \emph{Hotspot Guided} strategy, which directs drivers toward historically high-demand regions, improves upon random exploration but lacks real-time responsiveness to system dynamics, resulting in higher allocation times even when compared to the \emph{Greedy} approach. 
These results highlight the advantage of predictive, system-aware path planning in accelerating driver-passenger matching and improving operational efficiency.

\begin{table}[tb]
\caption{Startegy Performance Summary Table.}
\centering
\renewcommand{\arraystretch}{1.4}
\setlength{\tabcolsep}{4pt}
\small
\begin{tabular}{c|c|cccc}
\hline
\textbf{Fleet Size} & \textbf{Metric} & \textbf{WGC} & \textbf{Greedy} & \textbf{Random} & \textbf{Hotspot} \\
\hline
100  & Mean  & \textbf{15.91} & 16.07 & 19.78 & 20.05 \\
     & Worst & \textbf{47.84} & 53.06 & 50.74 & 56.58 \\
\hline
500  & Mean  & \textbf{32.42} & 36.96 & 45.76 & 42.84 \\
     & Worst & \textbf{75.89} & 84.51 & 100.44 & 104.58 \\
\hline
1000 & Mean  & \textbf{47.13} & 68.96 & 84.92 & 80.74 \\
     & Worst & \textbf{102.14} & 141.29 & 151.54 & 169.81 \\
\hline
2000 & Mean  & \textbf{64.85} & 112.97 & 145.91 & 135.76 \\
     & Worst & \textbf{162.68} & 231.28 & 223.28 & 236.26 \\
\hline
4000 & Mean  & \textbf{112.46} & 178.20 & 211.83 & 205.86 \\
     & Worst & \textbf{217.69} & 307.56 & 281.19 & 305.74 \\
\hline
5000 & Mean  & \textbf{119.63} & 212.67 & 232.09 & 231.18 \\
     & Worst & \textbf{226.87} & 333.31 & 288.72 & 311.08 \\
\hline
\end{tabular}
\label{tab:strategy_comparison}
\end{table}

\subsection{WGC Complexity Analysis}

The computational complexity of the WGC algorithm arises from two primary components: (i) numerical trajectory generation, and (ii) path enumeration with evaluation.

\paragraph{Numerical Trajectory Generation}
Let \( T \) denote the number of discrete time steps, \( E = |\mathcal{E}| \) the number of edges, \( V = |\mathcal{V}| \) the number of nodes, and \( \tau_{\max} \) the maximum edge travel time in discrete units. 
At each time step, WGC updates passenger queues, edge-level idle driver states, and node-level driver distributions via Euler integration. 
These updates also involve computing survival probabilities over a time-history window of length \( \tau_{\max} \).
The resulting complexity is:
\[
O\left(T \cdot (E \cdot \tau_{\max} + V)\right).
\]

\paragraph{Exhaustive Path Evaluation}
Starting from an initial node \( v_0 \), WGC exhaustively enumerates all simple paths up to a maximum length \( L \). 
In a graph with maximum out-degree \( d \), the number of such paths is \( O(d^L) \). For each path, the expected allocation time is computed via numerical integration along the trajectory, requiring \( O(\tau_{\max} \cdot L) \) operations. The overall complexity of this stage is thus
\[
O\left(d^L \cdot \tau_{\max} \cdot L\right),
\]
which is exponential in \( L \) and becomes computationally prohibitive even for moderate values of \( d \) and \( L \). This makes the overall computational complexity of WGC as:
\[
O\left(T \cdot (E \cdot \tau_{\max} + V) + d^L \cdot \tau_{\max} \cdot L\right).
\]

In practice, a typical overall computation time a 10x10 grid is approximately 15 sec.

\paragraph{Beam Search Acceleration.}
To overcome this bottleneck, we introduce a beam search strategy that retains only the top \( k \) partial paths at each expansion step, based on intermediate allocation time estimates. At each depth, each of the \( k \) candidates expands to at most \( d \) successors and is evaluated over a finite horizon of \( \tau_{\max} \cdot L \) time. This reduces the total complexity to
\[
O\left(k \cdot L \cdot d \cdot \tau_{\max}\right),
\]
which is linear in path length and graph branching factor, enabling real-time decision-making with negligible loss in path optimality. 

\section{Conclusion}
In this work, we introduced the \textit{Wise Goose Chase} (WGC) algorithm, a novel anticipatory path planning framework for idle drivers in ride-hailing systems.
Unlike conventional destination-based rebalancing strategies, WGC predicts the spatio-temporal dynamics of supply and demand across the road network and computes personalized cruising trajectories that minimize each driver's expected allocation time.
By explicitly modeling en-route matching opportunities and driver competition dynamics through a system of coupled delay differential equations, WGC enables more adaptive, context-aware guidance.

Extensive Monte Carlo simulations on synthetic urban networks show that WGC consistently outperforms standard baselines, including random walk, hotspot-guided, and static-demand greedy policies, across a range of fleet sizes and demand patterns.
These results highlight the importance of anticipatory, event-driven strategies over reactive heuristics in dynamic environments.

Future work will extend WGC to multi-agent coordination scenarios, where collective driver behavior must be jointly optimized to mitigate emergend inefficiencies.
We also plan to study the role of fairness and transparency in recommendation delivery, particularly how unequal or inaccurate guidance may erode driver trust, reduce compliance, and ultimately impact system-wide performance.

\bibliographystyle{IEEEtran}
\bibliography{ref.bib}

\begin{thebibliography}{10}
\providecommand{\url}[1]{#1}
\csname url@samestyle\endcsname
\providecommand{\newblock}{\relax}
\providecommand{\bibinfo}[2]{#2}
\providecommand{\BIBentrySTDinterwordspacing}{\spaceskip=0pt\relax}
\providecommand{\BIBentryALTinterwordstretchfactor}{4}
\providecommand{\BIBentryALTinterwordspacing}{\spaceskip=\fontdimen2\font plus
\BIBentryALTinterwordstretchfactor\fontdimen3\font minus \fontdimen4\font\relax}
\providecommand{\BIBforeignlanguage}[2]{{%
\expandafter\ifx\csname l@#1\endcsname\relax
\typeout{** WARNING: IEEEtran.bst: No hyphenation pattern has been}%
\typeout{** loaded for the language `#1'. Using the pattern for}%
\typeout{** the default language instead.}%
\else
\language=\csname l@#1\endcsname
\fi
#2}}
\providecommand{\BIBdecl}{\relax}
\BIBdecl

\bibitem{chen2021hunting}
C.~Chen, D.~Zhang, Y.~Wang, H.~Huang, C.~Chen, D.~Zhang, Y.~Wang, and H.~Huang, ``Hunting or waiting: Earning more by understanding taxi service strategies,'' \emph{Enabling Smart Urban Services with GPS Trajectory Data}, pp. 71--94, 2021.

\bibitem{pavone2012robotic}
M.~Pavone, S.~L. Smith, E.~Frazzoli, and D.~Rus, ``Robotic load balancing for mobility-on-demand systems,'' \emph{The International Journal of Robotics Research}, vol.~31, no.~7, pp. 839--854, 2012.

\bibitem{wallar2018vehicle}
A.~Wallar, M.~Van Der~Zee, J.~Alonso-Mora, and D.~Rus, ``Vehicle rebalancing for mobility-on-demand systems with ride-sharing,'' in \emph{2018 IEEE/RSJ international conference on intelligent robots and systems (IROS)}.\hskip 1em plus 0.5em minus 0.4em\relax IEEE, 2018, pp. 4539--4546.

\bibitem{brar2020ensuring}
A.~S. Brar and R.~Su, ``Ensuring service fairness in taxi fleet management,'' in \emph{2020 IEEE 23rd International Conference on Intelligent Transportation Systems (ITSC)}.\hskip 1em plus 0.5em minus 0.4em\relax IEEE, 2020, pp. 1--6.

\bibitem{sadeghi2019re}
A.~Sadeghi and S.~L. Smith, ``On re-balancing self-interested agents in ride-sourcing transportation networks,'' in \emph{2019 IEEE 58th Conference on Decision and Control (CDC)}.\hskip 1em plus 0.5em minus 0.4em\relax IEEE, 2019, pp. 5119--5125.

\bibitem{ong2021driver}
H.~Y. Ong, D.~Freund, and D.~Crapis, ``Driver positioning and incentive budgeting with an escrow mechanism for ride-sharing platforms,'' \emph{INFORMS Journal on Applied Analytics}, vol.~51, no.~5, pp. 373--390, 2021.

\bibitem{ZardiniAnnRev2022}
G.~Zardini, N.~Lanzetti, M.~Pavone, and E.~Frazzoli, ``Analysis and control of autonomous mobility-on-demand systems,'' \emph{Annual Review of Control, Robotics, and Autonomous Systems}, vol.~5, no.~1, pp. 633--658, 2022.

\bibitem{brar2024vehiclerebalancingadherenceuncertainty}
\BIBentryALTinterwordspacing
A.~S. Brar, R.~Su, and G.~Zardini, ``Vehicle rebalancing under adherence uncertainty,'' 2024. [Online]. Available: \url{https://arxiv.org/abs/2412.16632}
\BIBentrySTDinterwordspacing

\bibitem{chen2024rebalance}
H.~Chen, P.~Sun, Q.~Song, W.~Wang, W.~Wu, W.~Zhang, G.~Gao, and Y.~Lyu, ``i-rebalance: Personalized vehicle repositioning for supply demand balance,'' in \emph{Proceedings of the AAAI Conference on Artificial Intelligence}, vol.~38, no.~1, 2024, pp. 46--54.

\bibitem{guo2023vehicle}
G.~Guo, M.~Kang, and T.~Sun, ``Vehicle/employee rebalancing and charging scheduling in one-way car sharing systems,'' \emph{IEEE Transactions on Intelligent Transportation Systems}, vol.~24, no.~10, pp. 10\,665--10\,675, 2023.

\bibitem{brar2022supply}
A.~S. Brar, P.~Kasture, and R.~Su, ``Supply-demand balancing model for ev rental fleet,'' in \emph{2022 IEEE 25th International Conference on Intelligent Transportation Systems (ITSC)}.\hskip 1em plus 0.5em minus 0.4em\relax IEEE, 2022, pp. 1350--1355.

\bibitem{brar2021dynamic}
A.~S. Brar and R.~Su, ``Dynamic supply-demand balancing policy for cmod fleet,'' in \emph{2021 IEEE International Intelligent Transportation Systems Conference (ITSC)}.\hskip 1em plus 0.5em minus 0.4em\relax IEEE, 2021, pp. 2435--2440.

\bibitem{garg2018route}
N.~Garg and S.~Ranu, ``Route recommendations for idle taxi drivers: Find me the shortest route to a customer!'' in \emph{Proceedings of the 24th ACM SIGKDD international conference on knowledge discovery \& data mining}, 2018, pp. 1425--1434.

\end{thebibliography}

\end{document}